\documentclass[aps,showpcs,amsmath,amssymb,longbibliography,twocolumn]{revtex4-1}
\usepackage{multirow}
\usepackage{epsfig}
\usepackage{tikz}
\usetikzlibrary{quantikz}
\usepackage{amsmath}
\usepackage{bm,natbib}
\usepackage{times}
\usepackage{graphicx}
\usepackage{color}
\usepackage{slashed}
\usepackage{graphicx}
\usepackage{multirow}
\usepackage{amsmath,natbib} 
\usetikzlibrary{positioning,shapes}
\usepackage{relsize} 
\usepackage[latin1]{inputenc}
\usepackage{hyperref}  
\def\bea{\begin{eqnarray}}
\def\eea{\end{eqnarray}}
\def\bean{\begin{equation*}}
\def\eean{\end{equation*}}

\begin{document} 
 
\title{Leptophobic dark photon interpretation of the $\eta^{\left(\prime\right)}\rightarrow\pi^0\left(\eta\right)\gamma\gamma$ puzzle}

\author{Yaroslav~Balytskyi}
\email{ybalytsk@uccs.edu}
\affiliation{University of Colorado, Colorado Springs, Colorado, 80918, USA}

\date{\today}

\begin{abstract}

The decays of $\eta$ and $\eta^\prime$ mesons provide unique opportunities for testing the properties of low energy Quantum Chromodynamics and for the search of new physics beyond the Standard Model. However, recent experimental results on the rare decays of $\eta^{\left(\prime\right)}\rightarrow\pi^0\left(\eta\right)\gamma\gamma$ cannot be self-consistently described by the combination of the Vector Meson Dominance and Linear Sigma Model employing the same set of parameters. We show that this tension can be attributed to the presence of a leptophobic dark photon $\mathcal{B}$, and find representative values of the parameters which provide consistent description of these three decays, simultaneously. Unlike existing strategies of Dalitz analysis searching for the bump at $m_\mathcal{B}$, we propose the usage of mismatch between these decays to constrain the parameters of the hypothetical dark photon.
\end{abstract}

\maketitle 

\section{Introduction}
Long-lived neutral mesons $\eta$ and $\eta^\prime$ are a unique laboratory to study the properties of low-energy Quantum Chromodynamics (QCD), and a possible mismatch between theoretical predictions and experimental results is a sensitive probe for potential new physics beyond the Standard Model (BSM). To meet this challenge, extensive experimental programs in various high-intensity frontier centers studying these decays are deployed~\cite{Gan:2020aco}.

The doubly radiative decay $\eta\rightarrow\pi^0\gamma\gamma$ is particularly interesting because it is an ideal laboratory for testing the chiral perturbation theory and its extensions. The decays $\eta^\prime\rightarrow\pi^0\gamma\gamma$ and $\eta^\prime\rightarrow\eta\gamma\gamma$ are also interesting since they complete the calculations on the $\eta\rightarrow\pi^0\gamma\gamma$ decay. 

Early experimental results are summarized by Landsberg ~\cite{Landsberg:1985}, and other results have broadened the research landscape as follows: the branching ratio of the $\eta\to\pi^0\gamma\gamma$ decay was measured by GAMS-2000 to be $\rm{BR}^{\eta\rightarrow\pi^0\gamma\gamma}=(7.1\pm 1.4)\times 10^{-4}$~\cite{Alde1984}, CrystalBall@AGS in 2005 reported the value
$\rm{BR}^{\eta\rightarrow\pi^0\gamma\gamma}=(3.5\pm 0.7\pm 0.6)\times 10^{-4}$~\cite{Prakhov2005}, 
and, in 2008, Prakhov \textit{et al.} reported that $\mbox{BR}^{\eta\rightarrow\pi^0\gamma\gamma}=(2.21\pm 0.24\pm 0.47)\times 10^{-4}$~\cite{Prakhov2008} along with the invariant mass spectrum of the outgoing photons. The result of an independent analysis of the latest CrystalBall data is $\rm{BR}^{\eta\rightarrow\pi^0\gamma\gamma}=(2.7\pm 0.9\pm 0.5)\times 10^{-4}$~\cite{Knecht2004}. The 2006 result of the KLOE collaboration~\cite{dimillo2006} based on a sample of $68\pm 23$ events $\rm{BR}^{\eta\rightarrow\pi^0\gamma\gamma}=(0.84\pm 0.27\pm 0.14)\times 10^{-4}$, is lower in comparison with the previous measurements. The A2 collaboration at the Mainz Microtron (MAMI) reported the decay width and branching ratio to be $\Gamma(\eta\to\pi^{0}\gamma\gamma)=(0.330\pm 0.030)$ eV and
$\rm{BR}^{\eta\rightarrow\pi^0\gamma\gamma}=(2.54\pm 0.27)\times 10^{-4}$, based on the analysis of $1.2\times 10^{3}$ $\eta\to\pi^{0}\gamma\gamma$ decays~\cite{nefkens2014new}.  The most recent particle data group (PDG) value is $\rm{BR}^{\eta\rightarrow\pi^0\gamma\gamma}=(2.56\pm 0.22)\times 10^{-4}$~\cite{PDG}. Although it is possible to match this experimental value with theoretical predictions by appropriate choice of parameters, it seems impossible to reconcile the experimental data of all brother decays simultaneously, which we describe below.

Regarding the $\eta^\prime\to\pi^0\gamma\gamma$ decay, its width--including the invariant mass spectrum of two out-coming photons--was reported by the the BESIII collaboration~\cite{ablikim2017observation}. The branching fraction was measured to be
$\rm{BR} = \left(3.20\pm 0.07\pm 0.23\right)\times 10^{-3}$ superseding an  upper limit $\rm{BR} < 8 \times 10^{-4}$ at $90\%$ CL set by the GAMS-2000~\cite{alde1987neutral}. Finally, the branching for the $\eta^\prime\to\eta\gamma\gamma$ decay was bounded as $\rm{BR} < 1.33 \times 10^{-4}$ at $90\%$ CL, again by the BESIII collaboration~\cite{ablikim2019search}. In addition, JLab Eta Factory experiment plans to measure various $\eta^{\left(\prime\right)}$ decays with the GlueX apparatus and an upgraded forward calorimeter to test QCD and to search for BSM physics upon completion of construction in 2023~\cite{JLab1,JLab2,JLab3,JLab4,JLab5}.

From the point of view of theoretical calculations, the $\eta\rightarrow\pi^0\gamma\gamma$ decay has been studied in several frameworks. Vector Meson Dominance (VMD)~\cite{oppo1967models,baracca1970general}, chiral perturbation
theory ($\chi$PT)~\cite{ametller1992chiral}, with the further inclusion of C-odd axial-vector resonances \cite{ko1993contributions,ko1995eta}, by the approaches based on the unitarization of the chiral amplitudes~\cite{oset2003eta,oset2008eta}, dispersive formalisms~\cite{danilkin2017theoretical}, in the early and extended Nambu-Jona-Lasinio model~\cite{belkov1995,bellucci1995,bijnens1995}, as well as in the approach based on quark-box diagrams~\cite{ng1993,nemoto1996}. Preliminary results on the $\eta^\prime\rightarrow\pi^0\gamma\gamma$ and $\eta^\prime\rightarrow\eta\gamma\gamma$ decays were presented in~\cite{escribano2012,jora2010,Balytskyi:2018pzb,Balytskyi:2018uxb}. 

Finally, a combined analysis of all three decays $\eta^{\left(\prime\right)} \rightarrow\pi^0\left(\eta\right)\gamma\gamma$ was recently performed by Escribano \textit{et al.}~\cite{escribano2020theoretical}. In this work, the explicit contributions of the intermediate vector and scalar mesons were considered by employing the VMD and linear sigma model (L$\sigma$M). The coupling constants were parameterized by the phenomenological quark-based model fitted to the $V\rightarrow P\gamma$ and $P\rightarrow V\gamma$ decays~\cite{Bramon:2000fr,Escribano:2020jdy}. However, Escribano \textit{et al.} suggests that the \textit{simultaneous} fit of three decays is not possible, ``While a satisfactory description of the shape of the $\eta\rightarrow\pi^0\gamma\gamma$ and $\eta^\prime\rightarrow\pi^0\gamma\gamma$ decay spectra is obtained, thus supporting the validity of the approach, the corresponding branching ratios cannot be reproduced simultaneously"~\cite{escribano2020theoretical}.

In other words, by adjusting the parameters accordingly, we can reproduce the experimental results on $\eta\rightarrow\pi^0\gamma\gamma$ \textit{or} $\eta^\prime\rightarrow\pi^0\gamma\gamma$ and $\eta^\prime\rightarrow\eta\gamma\gamma$ \textit{separately} but not all three decays simultaneously. This discrepancy $\sim 5\sigma$ may indicate the limitation of the VMD+L$\sigma$M framework itself, ``This
puzzle might be pointing toward potential limitations of our theoretical treatment or, perhaps, the need for more precise measurements for the $\eta\rightarrow\pi^0\gamma\gamma$ decay, as our approach seems to be capable of successfully predicting the experimental data for the other two processes without the need for manual adjustment of the numerical input."~\cite{escribano2020theoretical}.

We propose considering another possibility to reproduce the experimental results \textit{simultaneously}. Assuming that the VMD+L$\sigma$M framework is valid and that the experimental results on all three decays of $\eta^{\left(\prime\right)}\rightarrow\pi^0\left(\eta\right)\gamma\gamma$ are correct, we propose an interpretation of this puzzle by an inclusion of the leptophobic dark photon or $\mathcal{B}$ boson.

Meson studies have long been considered as a discovery window for new vector or scalar mediators in the {\rm MeV-GeV} mass range~\cite{nelson1989, fayet2006}, and several observational anomalies can be an indication favoring such a scenario. First, excesses in high-energy cosmic rays can be interpreted as a dark matter annihilation into $e^+e^-$~\cite{arkani2009,pospelov2009}. Second, the issues with galactic rotation curves and other small-scale structure observations can be explained by self-interacting dark matter ~\cite{tulin2018}. Third, there is a discrepancy between the muon $\left(g - 2\right)_\mu$ theoretical Standard Model prediction~\cite{Aoyama2020} and the combined Brookhaven~\cite{Brookhaven} and Fermilab~\cite{FermiLab1,FermiLab2} experimental results. Updated and improved experimental results from Fermilab~\cite{FermiLab3} and J-PARC~\cite{JPARC} can shed new light on $\left(g - 2\right)_\mu$. This is an additional window of opportunity for
the BSM physics searches~\cite{fayet2007,pospelov20091}. Finally, the observation of anomalous $e^+e^-$ resonance in the ${}^8$Be decay~\cite{Be1,Be2,Be3,Be4,Be5,Be6,Be7} may be a possible indication of a new light gauge boson~\cite{krasznahorkay2016}. 

However, in all the aforementioned discrepancies and anomalies, the new force mediators are assumed to have predominantly leptonic coupling. In contrast to the previous studies, in this paper we propose to use the mismatch $\sim 5\sigma$ between the decays $\eta\rightarrow\pi^0\gamma\gamma$ and $\eta^\prime\rightarrow\pi^0\left(\eta\right)\gamma\gamma$ to fit the parameters of the hypothetical $\mathcal{B}$ boson, predominantly coupled to quarks rather than leptons, opening up new possibilities to search for this kind of BSM physics.

The paper is organized as follows. In Section~\ref{BProperties}, we discuss the properties of the hypothetical $\mathcal{B}$ boson and the validity of the assumption of ``naturalness" to relate the couplings of $\mathcal{B}$ to leptons versus those to quarks. In Section~\ref{Framework}, we provide a theoretical framework of $\rm{VMD+L\sigma M}$ with the inclusion of the hypothetical $\mathcal{B}$ boson, and investigate numerically its effects on the $\eta\rightarrow\pi^0\gamma\gamma$ and $\eta^\prime\rightarrow\pi^0\gamma\gamma$ observables in Section~\ref{Numerical}. We show a significantly improved fit of the experimental data demonstrated in Fig.~\ref{GammaGamma}. In our considerations, we limit our attention to $\eta\rightarrow\pi^0\gamma\gamma$ and $\eta^\prime\rightarrow\pi^0\gamma\gamma$ since the spectrum of the two out-coming photons for the case $\eta^\prime\rightarrow\eta\gamma\gamma$ is not yet available. Finally, we conclude in Section~\ref{Conclusions}.

\section{Properties of hypothetical $\mathcal{B}$ boson}\label{BProperties}

The aforementioned anomalies involving a new light gauge boson suggest that it primarily interacts with leptons rather than quarks. Another possibility that could avoid the dark photon searches is a vector mediator that interacts mainly with quarks rather than leptons. The minimal model of this kind is $U\left(1\right)_\mathcal{B}$ gauge symmetry of the baryon number. This possibility was first proposed in 1955 by Lee ~\cite{lee1955}, and later widely discussed in the literature~\cite{pais1973,rajpoot1989,foot1989,he1990,carone1995,bailey1995,carone19951,aranda1998}.

The argument in support of this model is that baryon number symmetry may have relation to dark matter~\cite{perez2010,agashe2004,farrar2006,davoudiasl2010,graesser,duerr2014} and could provide an explanation of dark matter stability on cosmological time scales as it carries a conserved baryon number charge. Additionally, in this model, dark matter and regular baryons have a joint baryogenesis which may be an explanation for the similarities between baryon and dark matter cosmic densities~\cite{nussinov1985}. Finally, the $\rm{U}\left(1\right)_\mathcal{B}$ gauge symmetry may serve as a natural framework for the Peccei-Quinn proposal for the resolution of the strong $\rm{CP}$ problem~\cite{foot1989,duerr2018}.

The searches of the $\rm{U}\left(1\right)_\mathcal{B}$ depend on its mass, which is not known a priori. Below the $\rm{MeV}$ mass range, low-energy neutron scattering~\cite{barbieri1975,leeb1992,nesvizhevsky2008} and other experiments~\cite{adelberger2003} strongly constrain new baryonic forces. Conversely, direct searches at high-energy colliders above the $\rm{GeV}$ did not reveal them~\cite{barger1996,dobrescu2013}. Therefore, studies of $\eta$ and $\eta^\prime$ mesons provide an opportunity to cover the intermediate regime of $\rm{MeV}-\rm{GeV}$~\cite{Gan:2020aco}. Additionally, invisible decays of the $\rm{U}\left(1\right)_\mathcal{B}$ gauge bosons may exist, and this opportunity can be probed by the neutrino factories producing a beam of dark matter particles scattering in the downstream detector~\cite{batell2014} and at higher mass range of $\rm{TeV}$ by a search of semi-visible jets at the LHC~\cite{cohen2015}. Finally, it has recently been proposed to test such kind of mechanism by the non-diffuse sources of gravitational waves through compact-object mergers~\cite{berryman2021,berryman2022}. However, a more detailed understanding of the conventional three-body forces in neutron star structure would be needed to perform this task.

The interaction Lagrangian describing the interaction of the hypothetical $\mathcal{B}$ boson with quarks $q$ and leptons $l$ is given by~\cite{nelson1989,tulin2014}:

\begin{equation}\label{InteractionLagrangian}
\mathcal{L}_{int}=(\frac{1}{3}g_{\mathcal{B}} + \epsilon\cdot Q_q\cdot e)\cdot \bar{q}\gamma^{\mu}q\cdot\mathcal{B_{\mu}}- \epsilon\cdot e \cdot \bar{l}\gamma^{\mu} l \cdot \mathcal{B_{\mu}}
\end{equation}
In order to preserve the symmetries of low-energy QCD, namely the invariance under the charge conjugation $\rm{(C)}$, parity inversion $\rm{(P)}$, and $\rm{SU(3)}$, it is required to have the same quantum numbers as the $\omega$ meson $\rm{I^G(J^{PC})=0^-(1^{--})}$. It can potentially manifest itself as a resonance in rare decays including $\eta^{\left(\prime\right)}\rightarrow\pi^0\left(\eta\right)\gamma\gamma$~\cite{tulin2014}. The mass of the hypothetical $\mathcal{B}$ boson is estimated to be in the range of 140 ${\rm MeV}$ - 1 ${\rm GeV}$.

Since the properties of the hypothetical $\mathcal{B}$ boson should be similar to those of the $\omega$ meson, and VMD was successfully applied  for description of the decays in Eqn.~(\ref{SimilarDecays}), and VMD was subsequently employed to 
explore the properties of hypothetical $\mathcal{B}$ boson~\cite{tulin2014} as well.

\begin{equation}\label{SimilarDecays}
    \omega\rightarrow \pi^0\gamma,\pi^+\pi^-\left(\pi^0\right); \ \pi^0, \eta^{\left(\prime\right)} \rightarrow\gamma\gamma; \ \omega,\phi\rightarrow\eta\gamma
\end{equation}

VMD was first employed in 1960 when Sakurai successfully applied Yang-Mills theory to strong interactions~\cite{sakurai1960}. Next, Kroll, Lee and Zumino made VMD electromagnetic form factors compatible with gauge invariance~\cite{kroll1960}. For the computations of hadronic processes where $\mathcal{B}$ boson is involved, we employ the hidden local symmetry (HLS) framework for VMD ~\cite{bando1985,bando19851,bando1988,fujiwara1985} which provides a low-energy effective theory for description of pseudoscalar meson nonet $\left(\pi^0, \eta, \eta^\prime, K, \bar{K}\right)$ and the vector meson nonet $\left(\rho^0,\omega,\phi,K^*,\bar{K}^*\right)$ with the latter being treated as a gauge boson of a hidden $U\left(3\right)_V$ symmetry.

Here we briefly summarize the HLS-VMD framework. According to~\cite{fujiwara1985}, the processes involving regular vector mesons arise from a single vector-pseudoscalar-vector vertex with the coupling constant fixed by the anomaly. The mixing with $V$ includes the external gauge fields $\left(\gamma,\mathcal{B}\right)$. The Feynman rule for mixing $V - \gamma$ is proportional to $e \rm{Tr}\left[\bf{Q}\cdot\bf{T_V}\right]$, where $\bf{Q}=diag\left(\frac{2}{3},-\frac{1}{3},-\frac{1}{3}\right)$ is a quark charge matrix, and the $\rm{U}\left(3\right)$ generator for $V$ is denoted by $\bf{T_V}$. The processes involving $\mathcal{B}$ boson can be obtained from the corresponding SM ones in Eqn.~(\ref{SimilarDecays}) by the replacement in the corresponding matrix element~\cite{tulin2014}:

\begin{equation}
e \rm{Tr}\left[\bf{Q}\cdot\bf{T_V}\right] \rightarrow \frac{g_B}{3}\rm{Tr}\left[\bf{T_V}\right]
\end{equation}
For example, the ratio of the decay width $\eta^\prime\rightarrow\mathcal{B}\gamma$ to $\eta^\prime\rightarrow\gamma\gamma$ is given by: 

\begin{multline}
    \frac{\Gamma\left(\eta^\prime \rightarrow \mathcal{B}\gamma \right)}{\Gamma\left(\eta^\prime \rightarrow \gamma\gamma \right)} \approx 2\frac{\alpha_B}{\alpha_{em}}\left(1 - \frac{m^2_{\mathcal{B}}}{m^2_{\eta^\prime}}\right)^3  \left(\frac{\rm{Tr}\left[\bf{T_{\eta^\prime}\bf{Q}}\right]}{3\rm{Tr}\left[\bf{T_{\eta^\prime}\bf{Q}^2}\right]})\right)^2 \\ = \frac{2}{49}\frac{\alpha_B}{\alpha_{em}}\left(1 - \frac{m^2_{\mathcal{B}}}{m^2_{\eta^\prime}}\right)^3,
\end{multline}
where $\bf{T_{\eta^\prime}}$ represents the generator of $\bf{\eta^\prime}$ and flavor $\rm{SU}\left(3\right)$-breaking effects in the pseudoscalar form factors were neglected.

The lack of alternatives to the VMD assumption does not validate its application, and the need to critically assess it by taking into account the QCD constraints was shown in~\cite{ivanov2008,du2020}. The veracity of the $\rm{VMD}$ approach as a tool for exploring the relation between the electromagnetic vector-meson production reaction $e + p \rightarrow e^\prime + V + p$ and the purely hadronic process $V + p \rightarrow V + p$, taking into account the QCD constraints, was investigated in~\cite{xu2021}. This analysis suggests that the $\rm{VMD}$ approach may be reasonable for light vector mesons. However, in cases where vector mesons are described by the momentum-dependent bound-state amplitudes, $\rm{VMD}$ does not work for heavy vector mesons and is unable to reliably predict the photon-to-vector-meson transition strength or the momentum dependence of the integrands appearing in the calculations of the reaction amplitudes. Therefore, in application to the case of our interest, since the involved particles are light, the $\rm{VMD}$ should be reasonable when applied to $\eta^{\left(\prime\right)}\rightarrow\pi^0\left(\eta\right)\gamma\gamma$. Additionally, the $\rm{VMD}$ ideas are incorporated into the rigorous $\chi$PT Lagrangian~\cite{donoghue1989} with $\rm{VMD}$ giving a dominant contribution and the $\chi$PT corrections contributing several percent, as discussed in Section~\ref{Numerical}. In our numerical calculations in Section~\ref{Numerical}, we explicitly take these effects into account. 

Finally, we need to discuss the coupling constants in the interaction Lagrangian in Eqn.~(\ref{InteractionLagrangian}). Regarding the quark coupling $g_\mathcal{B}$, the requirement of ``naturalness" under the assumption that the masses of new fermions are generated by the $\rm{U}\left(1\right)_\mathcal{B}$-breaking Higgs field with $g_{\mathcal{B}}$ being the $\rm{U}\left(1\right)_\mathcal{B}$ gauge coupling, given by~\cite{williams2011}:

\begin{equation}\label{QuarkCoupling}
    g_{\mathcal{B}} \lesssim 10^{-2}\times \left(\frac{m_{\mathcal{B}}}{100 \rm{MeV}}\right)
\end{equation}

The coupling to quarks in the model described by the Lagrangian in Eqn.~(\ref{InteractionLagrangian}) should dominate the coupling to leptons. However, due to existence of the kinetic mixing between $\mathcal{B}$ and photon, $\mathcal{B}$ boson may not be completely decoupled from leptons and this effect is described by the $\epsilon$ parameter. In the case when $\epsilon$ vanishes at tree level, it can be generated to be $\epsilon\ne 0$ by the loop corrections involving heavy quarks, and a typical size of this correction is~\cite{carone19951,aranda1998}:

\begin{equation}\label{LeptonQuark}
\epsilon\sim e\frac{g_B}{\left(4\pi\right)^2}  
\end{equation}

In contrast~\cite{tulin2014} where the couplings of $\mathcal{B}$ boson to mesons are given assuming an exact $\rm{SU}\left(3\right)$ flavor symmetry, we consider them without the reference to it. Since $\mathcal{B}$ boson should have the same quantum numbers as $\omega$ meson, it should possess coupling analogous to $VP\gamma$, $g_{\mathcal{B}\rightarrow\eta\gamma}$, $g_{\mathcal{B}\rightarrow\pi^0\gamma}$, and $g_{\eta^\prime\rightarrow\mathcal{B}\gamma}$. Analogously to the couplings of vector mesons described in Section~\ref{Framework}, the couplings of $\mathcal{B}$ boson in the  $\eta^{\left(\prime\right)}\rightarrow\pi^0\left(\eta\right)\gamma\gamma$ decays are given by a product of the couplings, without any reference to the flavor $\rm{SU}\left(3\right)$ symmetry:

\begin{equation}\label{BCouplings}
\begin{cases}
c^\mathcal{B}_{\eta\rightarrow\pi^0\gamma\gamma} = g_{\mathcal{B}\rightarrow\eta\gamma} \cdot g_{\mathcal{B}\rightarrow\pi^0\gamma}\\ 
c^\mathcal{B}_{\eta^\prime\rightarrow\pi^0\gamma\gamma} = g_{\eta^\prime\rightarrow\mathcal{B}\gamma} \cdot g_{\mathcal{B}\rightarrow\pi^0\gamma} \\
c^\mathcal{B}_{\eta^\prime\rightarrow\eta\gamma\gamma} = g_{\eta^\prime\rightarrow\mathcal{B}\gamma} \cdot g_{\mathcal{B}\rightarrow\eta\gamma}
\end{cases}
\end{equation}
As we discuss in detail in Section~\ref{Numerical}, to describe self-consistently the $\eta^{\left(\prime\right)}\rightarrow\pi^0\left(\eta\right)\gamma\gamma$ decays, we require $m_{\mathcal{B}} > m_\eta$ since the mass of $\mathcal{B}$ boson is close to the mass of $\omega$ meson, as we discuss in Section~\ref{Numerical},  Eqn.~(\ref{condition}). The partial decay widths, assuming $m_{\mathcal{B}} > m_\eta$, are given by:

\begin{equation}
    \Gamma\left(\mathcal{B}\rightarrow\left(\pi^0\right)\eta\gamma\right) = \frac{g^2_{\mathcal{B}\rightarrow\left(\pi^0\right)\eta\gamma}}{3}\frac{\left(m^2_{\mathcal{B}} - m^2_{\left(\pi^0\right)\eta}\right)^3}{32\pi m^3_{\mathcal{B}}}
\end{equation}

\begin{equation}\label{Invisible}
    \Gamma\left(\eta^\prime\rightarrow\mathcal{B}\gamma\right) = g^2_{\eta^\prime\rightarrow\mathcal{B}\gamma}\frac{\left(m^2_{\eta^\prime} - m^2_{\mathcal{B}}\right)^3}{32\pi m^3_{\eta^\prime}}
\end{equation}
The coupling constant $g_{\eta^\prime\rightarrow\mathcal{B}\gamma}$ in Eqn.~(\ref{Invisible}) is limited by the bound on the invisible decays of $\eta^\prime$, given by PDG~\cite{PDG} as follows:
\begin{equation}
\rm{Br}\left(\eta^\prime\rightarrow invisible\right) < 6\times 10^{-4}    
\end{equation}
Similar constraints from the $\rm{SU(3)}$ symmetry can be obtained as shown in~\cite{tulin2014}.
Unlike the "bump-hunting" strategy proposed in~\cite{tulin2014}, by performing a fit of the couplings in Eqn.~(\ref{BCouplings}) and fitting to all three decays $\eta^{\left(\prime\right)}\rightarrow\pi^0\left(\eta\right)\gamma\gamma$, it is possible to determine the values of the couplings of $\mathcal{B}$ boson from an Eqn.~(\ref{BCouplings}) as:

\begin{equation}
g_{\mathcal{B}\rightarrow\eta\gamma} = \sqrt{\frac{c^\mathcal{B}_{\eta\rightarrow\pi^0\gamma\gamma}\cdot c^\mathcal{B}_{\eta^\prime\rightarrow\eta\gamma\gamma}}{c^\mathcal{B}_{\eta^\prime\rightarrow\pi^0\gamma\gamma}}}    
\end{equation}

\begin{equation}
g_{\mathcal{B}\rightarrow\pi^0\gamma} = \sqrt{\frac{c^\mathcal{B}_{\eta^\prime\rightarrow\pi^0\gamma\gamma}\cdot c^\mathcal{B}_{\eta\rightarrow\pi^0\gamma\gamma} }{c^\mathcal{B}_{\eta^\prime\rightarrow\eta\gamma\gamma}}}    
\end{equation}

\begin{equation}
g_{\eta^\prime\rightarrow\mathcal{B}\gamma} = \sqrt{\frac{c^\mathcal{B}_{\eta^\prime\rightarrow\eta\gamma\gamma}\cdot c^\mathcal{B}_{\eta^\prime\rightarrow\pi^0\gamma\gamma} }{c^\mathcal{B}_{\eta\rightarrow\pi^0\gamma\gamma}}}   \end{equation}

As we show in Section~\ref{Numerical} in Table~\rm{II}, in order to explain the discrepancies in  $\eta^{\left(\prime\right)}\rightarrow\pi^0\left(\eta\right)\gamma\gamma$, we need to require that the couplings of $\mathcal{B}$ boson to $\eta$ be relatively large and of the same order of magnitude to that of $\omega$:
\begin{equation}\label{OrderOfMagnitude}
c^{\mathcal{B}}_{\eta\rightarrow\pi^0\gamma\gamma} = g_{\mathcal{B}\pi^0\gamma} \cdot g_{\mathcal{B}\eta\gamma} \sim c^{\omega}_{\eta\rightarrow\pi^0\gamma\gamma} = g_{\omega\pi^0\gamma} \cdot g_{\omega\eta\gamma}    
\end{equation}
At the same time, the experimental results by BaBar~\cite{babar2014,babar2016,babar2017,babar2022} and KLOE-2~\cite{kloe2012,kloe2016,kloe2018} put a stringent constraint on the leptonic coupling to be $\epsilon< 10^{-3} - 10^{-5}$. The ``naturalness" requirements in Eqns.~(\ref{QuarkCoupling}) and (\ref{LeptonQuark}), which expect the coupling to be small, seem to be in dissension with Eqn.~(\ref{OrderOfMagnitude}). 

Nevertheless, the ``naturalness" and ``fine-tuning" requirements are in question in current work,~\cite{altarelli2014,hossenfelder2021}. Therefore, in order to explain the 
$\eta^{\left(\prime\right)}\rightarrow\pi^0\left(\eta\right)\gamma\gamma$ mismatch, the ``naturalness" requirement in Eqns.~(\ref{QuarkCoupling}) and (\ref{LeptonQuark}) should instead be replaced with:

\begin{equation}\label{Inverse}
\epsilon\ll e\frac{g_B}{\left(4\pi\right)^2}    
\end{equation}
In other words, the ``unnatural" values of the coupling constants of $\mathcal{B}$ boson are required in order to explain the experimental data on the $\eta^{\left(\prime\right)}\rightarrow\pi^0\left(\eta\right)\gamma\gamma$ decays.

\section{Theoretical framework}\label{Framework}

$\eta$ and $\eta^\prime$ are hadrons, and thus the description of their properties inevitably involves strong interactions described by Quantum Chromodynamics (QCD). However, the QCD coupling constant is large at low energies and, as a result, the perturbative description of these processes in terms of quarks and gluons is not possible. The only truly {\it ab initio} approach to deal with non-perturbative QCD processes at low energies is lattice QCD (reviewed here~\cite{Gan:2020aco}) alsong with a number of recent advances for $\eta$ and $\eta^\prime$ mesons which were done by~\cite{bali2021}. However, lattice QCD is still currently unable to describe the underlying dynamics of the $\eta$ and $\eta^\prime$ mesons completely, and other methods need to be employed.

As discussed in the previous section, the dominant contribution in the decays which we consider are given by the VMD. However, it has to be incorporated into a rigorous $\chi$PT framework pioneered by Weinberg~\cite{weinberg1979} and Gasser and Leutwyler ~\cite{gasser1984,gasser1985}, which we describe below. Since at low energies the perturbative description of QCD is not possible, another approach, $\chi$PT, which is an effective field theory based on chiral symmetry,  provides an expansion on momenta of the involved particles $p^2$ and has proven itself to be a powerful tool. In this theory, low-energy constants need to be fixed from the observables and it often needs to be complemented with dispersion theory to reach the required precision. The extension of this theory, called resonance chiral theory (R$\chi$T)~\cite{RChTReview}, may shed new light on the phenomena involving $\eta-\eta^\prime$ mesons and require additional theoretical efforts.

The $\eta\rightarrow\pi^0\gamma\gamma$ decay is considered a rigorous test for the predictive power of $\chi$PT as described in this seminal work~\cite{ametller1992chiral}. Since the involved pseudoscalar mesons are neutral, the tree-level contributions at $O\left(p^2\right)$ and $O\left(p^4\right)$ vanish. The first nonzero contribution comes in at $O\left(p^4\right)$ from the kaon and pion loops, and the latter are suppressed due to violation of $\rm{G}$-parity and thus proportional to $m_u - m_d$. The only sizable contribution in this decay comes at $O\left(p^6\right)$. To fix the associated low-energy constants, VMD was used~\cite{ametller1992chiral}, and the corresponding constants were fixed by expanding the vector meson propagators in powers of $t/M^2_V$ and $u/M^2_V$. The $O\left(p^8\right)$ loop corrections with two anomalous vertices are negligibly small~\cite{ametller1992chiral}.

In our work, we adopt the approach developed in~\cite{escribano2020theoretical}, and implement the $\mathcal{B}$ boson effects by modifying the VMD part of the amplitude. In this approach, the large-$N_c$ and isospin limits are assumed, and the singlet state $\eta_0$ is treated as the ninth pseudo-Goldstone boson of the theory. As a result, for $\eta^{\left(\prime\right)}\rightarrow\pi^0\gamma\gamma$ only kaon loop is involved. The effects of vector mesons decaying through the chain $\eta^{\left(\prime\right)}\rightarrow V\gamma \rightarrow \pi^0\gamma\gamma$ are accounted for by VMD, and the $\rm{L\sigma  M}$ explicitly takes into account the effects of scalar meson resonances. Scalar meson poles can be included at the same time as keeping the correct low-energy properties expected from chiral symmetry by using the complementarity between $\rm{L\sigma M}$ model and $\chi$PT. Such procedure was successfully used for the $V \rightarrow P^0 P^0 \gamma$ processes~\cite{escribano2006}. The loop corrections arising from the diagrams with two anomalous vertices are negligible and are thus neglected. 

As was shown in~\cite{escribano2020theoretical}, for the case of $\eta\rightarrow\pi^0\gamma\gamma$,  vector mesons give $\approx 93\%$ of the total decay width, and the remaining $\approx 7\%$ correspond to their {\it constructive} interference with the scalar mesons, and for the $\eta^\prime\rightarrow\pi^0\gamma\gamma$ case, the corresponding interference is $\approx -0.4\%$ and vector mesons completely dominate giving $\approx 100\%$ of the total decay width.

The total matrix element is obtained as a coherent sum of VMD+L$\sigma$M:

\begin{equation}
    \lvert \mathcal{M}\lvert^2 =   \lvert \mathcal{M}^{\rm{VMD}}\lvert^2   +  2\rm{Re} \left(\mathcal{M}^{\rm{VMD}}\left(\mathcal{M}^{L \sigma M}\right)^\dagger\right)+\lvert \mathcal{M}^{L \sigma M}\lvert^2.
\end{equation}
Similarly to~\cite{escribano2020theoretical}, in our calculation, VMD and L$\sigma$M contributions are taken with no relative phase. However, for the case of $\eta\rightarrow\pi^0\gamma\gamma$ where the interference is significant, even if included, it is unable to explain the discrepancy with the experimental result since it is {\it constructive}, and the theoretical prediction is $\sim 2$ times {\it less} than the experimental result. Therefore, $\sim 7\%$ contribution from scalar and vector meson interference is insufficient to account for it.

The VMD part of the total matrix element for the $\eta^{\left(\prime\right)}\rightarrow\pi^0\left(\eta\right)\gamma\gamma$ decays is determined by the Eqn.~(\ref{VMD}):

\begin{widetext}
\begin{eqnarray}\label{VMD}
\quad {\cal M}^{\mathrm{VMD}}_{\eta^{\left(\prime\right)}\to\pi^0\left(\eta\right)\gamma\gamma}=
\sum_{V=\rho^0, \omega, \phi}g_{V\!\eta^{\left(\prime\right)}\gamma}g_{V\!\pi^0\left(\eta\right)\gamma}\left[\frac{(P\cdot q_2-m_{\eta^{\left(\prime\right)}}^2)\{a\}-\{b\}}{D_V(t)}+
\bigg\{
\begin{array}{c}
q_2\leftrightarrow q_1\\
t\leftrightarrow u
\end{array}
\bigg\}\right]\ ,
\end{eqnarray}
\end{widetext}
with $t,u=(P-q_{2,1})^2=m^2_{\eta^{\left(\prime\right)}}-2P\cdot q_{2,1}$ being the Mandelstam variables. In our conventions, the particles are numbered as $\{1,2,3\}  = \{\gamma,\gamma,\pi^0\}$. The Lorentz structures $\{a\}$ and $\{b\}$ are given by:

\begin{equation}
\begin{aligned}
&\{a\} = (\epsilon_1\cdot\epsilon_2)(q_1\cdot q_2)-(\epsilon_1\cdot q_2)(\epsilon_2\cdot q_1) \ , \\
&\{b\} =(\epsilon_1\cdot q_2)(\epsilon_2\cdot P)(P\cdot q_1)+(\epsilon_2\cdot q_1)(\epsilon_1\cdot P)(P\cdot q_2) -\\
&-(\epsilon_1\cdot\epsilon_2)(P\cdot q_1)(P\cdot q_2)-(\epsilon_1\cdot P)(\epsilon_2\cdot P)(q_1\cdot q_2), \ 
\end{aligned}
\end{equation}
where $P$ is the four-momentum of the decaying $\eta\left(\eta^\prime\right)$, $\epsilon_{1,2}$ and $q_{1,2}$ are the polarisation and four-momentum of the final photons, respectively. The propagator of the vector meson is given by:

\begin{equation}\label{BW}
D_V(t)=m_V^2-t-i\,m_V\Gamma_V,
\end{equation}
where $V=\omega$, $\rho^0$, and $\phi$. 
The widths of $\omega$ and $\phi$ mesons are much smaller than that of $\rho^0$. Therefore, in our calculations, the decay widths of the $\omega$ and $\phi$ mesons are kept constant, whereas for the $\rho^0$ meson we employ an energy-dependent parameterization to take it into account. Unlike~\cite{escribano2020theoretical}, in our calculations we apply a new parameterization of the $\rho^0$-meson decay width~\cite{lichard2006}, instead of~\cite{roos1969} previously used in~\cite{escribano2020theoretical}:
\begin{equation}
\Gamma_{\rho^0}(s)=\Gamma_{\rho^0}\frac{m_{\rho^0}}{\sqrt{s}}\left(\frac{s-4  m^2_{\pi^+}}{m^2_{\rho^0}-4  m^2_{\pi^+}}\right)^{\frac{3}{2}} \theta\left(s-4 m^2_{\pi^+}\right),
\end{equation}
because it provides equally good or even better fits to the CMD2, SND, and KLOE data~\cite{lichard2006}. We neglect possible effects of mixing $\mathcal{B}$ boson with $\rho^0$ meson and possible modification of the $\rho^0$ decay width due to this since $\rho^0$ meson is isovector while $\mathcal{B}$ boson is isoscalar.

In case one assumes an exact $\rm{SU}\left(3\right)$-flavour symmetry and an OZI-rule limit, it is possible to express all $g_{VP\gamma}$ couplings by a single coupling constant and $\rm{SU}\left(3\right)$-group factors~\cite{bramon1995}. To take into account the effects of $\rm{SU}\left(3\right)$-flavour symmetry breaking and an OZI-rule violation--which are inevitably present---we employ a phenomenological model~\cite{Bramon:2000fr,Escribano:2020jdy} which was previously developed for the description of $V\rightarrow P\gamma$ and $P \rightarrow V \gamma$ decays. In this model, the differences in the effective magnetic moments of the light $u$, and $d$, and strange quarks in magnetic dipolar transitions, which correspond to the flavor symmetry breaking, are represented by constituent quark mass differences by an introduction of a multiplicative $\rm{SU}\left(3\right)$-breaking factor $1 - s_e \equiv \frac{\bar{m}}{m_s}$ in the $s$-quark component of the quark charge matrix ${\bf Q}$. The corresponding coupling constants are given by:

\begin{equation}
\begin{aligned}
& g_{\rho^0\pi^0\gamma} = \frac{g}{3}, \ g_{\rho^0\eta\gamma} = gz_{\textrm{NS}}\cos{\varphi_P}, \ g_{\rho^0\eta^{\prime}\gamma} = gz_{\textrm{NS}}\sin{\varphi_P},
\\ 
& g_{\omega \pi^0 \gamma} = g\cos{\varphi_V}, \\
& g_{\omega\eta\gamma} = \frac{g}{3}\left(z_{\textrm{NS}}\cos{\varphi_P}\cos{\varphi_V} - 2 \frac{\overline{m}}{m_s}z_{\rm{S}}\sin{\varphi_P}\sin{\varphi_V}\right), \\
& g_{\omega\eta^{\prime}\gamma} = \frac{g}{3}\left(z_{\textrm{NS}}\sin{\varphi_P}\cos{\varphi_V} + 2 \frac{\overline{m}}{m_s}z_{\rm{S}}\cos{\varphi_P}\sin{\varphi_V}\right), \\
& g_{\phi\pi^0\gamma} = g\sin{\varphi_V}, \\
& g_{\phi\eta\gamma} = \frac{g}{3}\left(z_{\rm{NS}}\cos{\varphi_P}\sin{\varphi_V} + 2 \frac{\overline{m}}{m_s}z_{\rm{S}}\sin{\varphi_P}\cos{\varphi_V}\right), \\
& g_{\phi\eta^{\prime}\gamma} = \frac{g}{3}\left(z_{\rm{NS}}\sin{\varphi_P}\sin{\varphi_V} - 2 \frac{\overline{m}}{m_s}z_{\rm{S}}\cos{\varphi_P}\cos{\varphi_V}\right)\ ,
\end{aligned}
\end{equation}
here $g$ represents a generic electromagnetic constant, $\varphi_P$ is a pseudoscalar $\eta-\eta^\prime$ mixing angle in the quark-flavor basis, which, at lowest order in $\chi$PT, is defined as:

\begin{equation}
\begin{cases}
\ket{\eta} = \ket{\eta_{NS}}\cos\left(\varphi_P\right) - \ket{\eta_S}\sin\left(\varphi_P\right) \\
\ket{\eta^\prime} = \ket{\eta_{NS}}\sin\left(\varphi_P\right) + \ket{\eta_S}\cos\left(\varphi_P\right) 
\end{cases},
\end{equation}
where $\ket{\eta_{NS}} = \frac{1}{\sqrt{2}}\left(\ket{u\bar{u}} + \ket{d\bar{d}}\right)$ and $\ket{\eta_S} = \ket{s\bar{s}}$~\cite{bramon1997}. $\varphi_V$ represents the vector $\omega-\phi$ mixing angle in the basis defined above, while $\bar{m}/m_S$ is the ratio of constituent quark masses, and the relative meson wave-function overlaps are accounted for by the non-strange and strange multiplicative factors given $z_{NS}$ and $z_S$, respectively.

The result of the fit $\# 4$~\cite{Escribano:2020jdy} was previously used in~\cite{escribano2020theoretical} and is given by:

\begin{equation}\label{fit4}
\begin{aligned}
& g = 0.70 \pm 0.01 \ \rm{GeV}^{-1},	\varphi_{P} = (41.4 \pm 0.5)^\circ, \\
& z_{\rm{NS}} = 0.83 \pm 0.02, \	z_{\rm{S}}\overline{m}/m_s = 0.65 \pm 0.01, \\ & \varphi_{V} = (3.3 \pm 0.1)^\circ,
\end{aligned}
\end{equation}
This fit provides good agreement with the experimental results on $V\rightarrow P\gamma$ and $P\rightarrow V\gamma$ illustrated below~\cite{Escribano:2020jdy}:
\begin{table}
\begin{center}
\begin{tabular}{|c|c|c|} 
\hline
Decay  & $\Gamma_{exp.} \left(keV\right)$ & $\Gamma_{th.} \left(keV\right)$ \\
\hline
$\rho^0\rightarrow\eta\gamma$ &  $44 \pm 3$ & $38 \pm 2$ \\ \hline
$\rho^0\rightarrow\pi^0\gamma$ &  $69 \pm 9$ & $79 \pm 2$  \\ 
\hline
$\omega\rightarrow\eta\gamma$ & $3.8 \pm 0.3$ & $3.5 \pm 0.2$   \\
\hline
$\omega\rightarrow\pi^0\gamma$ & $713 \pm 20$ & $704 \pm 19$   \\
\hline
$\phi\rightarrow\eta\gamma$ & $54.4 \pm 1.1$ & $54 \pm 8$   \\
\hline
$\phi\rightarrow\eta^\prime\gamma$ & $0.26 \pm 0.01$ & $0.27 \pm 0.05$   \\
\hline
$\phi\rightarrow\pi^0\gamma$ & $5.5 \pm 0.2$ & $ 5.5 \pm 0.3$   \\
\hline
$\eta^\prime\rightarrow\rho^0\gamma$ & $57 \pm 3$ & $55 \pm 3$   \\
\hline
$\eta^\prime\rightarrow\omega\gamma$ & $5.1 \pm 0.3$ & $6.5 \pm 0.1$   \\ \hline
\end{tabular}
\caption{$V\rightarrow P\gamma$ and $P\rightarrow V\gamma$ decays with the parameters from Eqn.~(\ref{fit4}).}
\end{center}
\end{table}

The L$\sigma$M part of the amplitude has the form:
\begin{equation}
{\cal M}^{\mathrm{L\sigma M}}_{\eta^{\left(\prime\right)}\to\pi^0\gamma\gamma} =
\frac{2\alpha}{\pi}\frac{1}{m_{K^+}^2}L(s_K)\{a\}\times{\cal M}^{\rm{L\sigma M}}_{K^+K^-\to\pi^0\eta^{\left(\prime\right)}},
\end{equation}
where the loop integrals are given by:
\begin{equation}
\begin{aligned}
&L\left(z\right)=-\frac{1}{2z}-\frac{2}{z^2}f\left(\frac{1}{z}\right),\\
&f\left(z\right)=
\begin{cases}{}
\frac{1}{4}\left(\log\frac{1+\sqrt{1-4z}}{1-\sqrt{1-4z}}-i\pi\right)^2, & \mbox{if}\ z<\frac{1}{4}\\[1ex]
-\left[\arcsin\left(\frac{1}{2\sqrt{z}}\right)\right]^2, & \mbox{if}\ z>\frac{1}{4}
\end{cases},
\end{aligned}
\end{equation}
and $s_K=s/m_{K^+}^2$, $s=(q_1+q_2)^2=2q_1\cdot q_2$ is the invariant mass of the outgoing photons.

Defining $\beta_K=\sqrt{1-4m_K^2/s}$,
$\bar\beta_K=\sqrt{4m_K^2/s-1}$, $\theta_K=\theta(s-4m_K^2)$, and 
$\bar\theta_K=\theta(4m_K^2-s)$, $\beta^\pm_{\pi\eta^{\left(\prime\right)}}=\sqrt{1-(m_\pi\pm m_{\eta^{\left(\prime\right)}})^2/s}$,
$\bar\beta^\pm_{\pi\eta^{\left(\prime\right)}}=\sqrt{(m_\pi\pm m_{\eta^{\left(\prime\right)}})^2/s-1}$,
$\theta_{\pi\eta^{\left(\prime\right)}}=\theta[s-(m_\pi+m_{\eta^{\left(\prime\right)}})^2]$,
$\bar\theta_{\pi\eta^{\left(\prime\right)}}=\theta[s-(m_\pi-m_{\eta^{\left(\prime\right)}})^2]\times\theta[(m_\pi+m_{\eta^{\left(\prime\right)}})^2-s]$, and 
$\bar{\bar\theta}_{\pi\eta^{\left(\prime\right)}}=\theta[(m_\pi-m_{\eta^{\left(\prime\right)}})^2-s]$, the propagator of the scalar meson takes the form:

\begin{equation}
D(s)=s-m_R^2+{\rm Re}\Pi(s)-{\rm Re}\Pi(m_R^2)+ i{\rm Im}\Pi(s),\
\end{equation}
where ${\rm Re}\Pi(s)$ and ${\rm Im}\Pi(s)$ are given in Eqns.~(\ref{Rea0}) and~(\ref{Ima0}).

\begin{widetext}
\begin{equation}
\begin{aligned}
\label{Rea0}
\qquad\qquad\qquad {\rm Re}\left(\Pi(s)\right)=&\frac{g_{a_0 K\bar K}^2}{16\pi^2}
\left[2-\beta_K\log\left(\frac{1+\beta_K}{1-\beta_K}\right)\theta_K
-2\bar\beta_K\arctan\left(\frac{1}{\bar\beta_K}\right)\bar\theta_K\right] + \\
&+\frac{g_{a_0\pi\eta^{\left(\prime\right)}}^2}{16\pi^2}
\Bigg[2-\frac{m^2_{\eta^{\left(\prime\right)}}-m^2_\pi}{s}\log\left(\frac{m_{\eta^{\left(\prime\right)}}}{m_\pi}\right)
-\beta^+_{\pi\eta^{\left(\prime\right)}}\beta^-_{\pi\eta^{\left(\prime\right)}}
\log\left(\frac{\beta^-_{\pi\eta^{\left(\prime\right)}}+\beta^+_{\pi\eta^{\left(\prime\right)}}}{\beta^-_{\pi\eta^{\left(\prime\right)}}-\beta^+_{\pi\eta^{\left(\prime\right)}}}\right)
\theta_{\pi\eta^{\left(\prime\right)}} \\
&-2\bar\beta^+_{\pi\eta^{\left(\prime\right)}}\beta^-_{\pi\eta^{\left(\prime\right)}}\arctan\left(\frac{\beta^-_{\pi\eta^{\left(\prime\right)}}}{\bar\beta^+_{\pi\eta^{\left(\prime\right)}}}\right)
\bar\theta_{\pi\eta^{\left(\prime\right)}}
+\bar\beta^+_{\pi\eta^{\left(\prime\right)}}\bar\beta^-_{\pi\eta^{\left(\prime\right)}}
\log\left(\frac{\bar\beta^+_{\pi\eta^{\left(\prime\right)}}+\bar\beta^-_{\pi\eta^{\left(\prime\right)}}}{\bar\beta^+_{\pi\eta^{\left(\prime\right)}}-\bar\beta^-_{\pi\eta^{\left(\prime\right)}}}\right)
\bar{\bar\theta}_{\pi\eta^{\left(\prime\right)}} \Bigg] ,
\end{aligned} 
\end{equation}\\[-6ex]
\begin{equation}
\begin{aligned}
\label{Ima0}
\qquad\qquad\qquad {\rm Im}\left(\Pi(s)\right)=-\frac{g_{a_0 K\bar K}^2}{16\pi}\beta_K\theta_K
      -\frac{g_{a_0\pi\eta}^2}{16\pi}\beta^+_{\pi\eta}\beta^-_{\pi\eta}\theta_{\pi\eta}.
\end{aligned}
\end{equation}
\end{widetext}
The couplings of the $a_0$ to kaons in the isospin limit are~\cite{escribano2020theoretical}:

\begin{equation}
\begin{cases}
g_{a_0 K\bar K}^2=2g_{a_0 K^+K^-}^2=\frac{1}{2}\left(\frac{m_{K}^2-m_{a_0}^2}{f_{K}}\right)^2 \\
g_{a_0 \pi \eta}^2=\left(\frac{m_{\eta}^2-m_{a_0}^2}{f_{\pi}}\cos{\varphi_P}\right)^2 \\
g_{a_0 \pi \eta^\prime}^2=\left(\frac{m_{\eta^\prime}^2-m_{a_0}^2}{f_{\pi}}\sin{\varphi_P}\right)^2
\end{cases}
\end{equation}

In our numerical calculations, the renormalized mass of the $a_0$ meson is taken as $m_{a_0}=980$ {\rm MeV}, and $f_{\pi} = 92.07$ {\rm MeV}, $f_K = 110.10$ {\rm MeV}.

These input values of the parameters for the VMD+L$\sigma$M described above provide reasonable agreement with the experimental data on the $\eta^\prime\rightarrow\pi^0\gamma\gamma$ and $\eta^\prime\rightarrow\eta\gamma\gamma$ decays, but lead to a $\sim 5\sigma$ mismatch for the $\eta\rightarrow\pi^0\gamma\gamma$ decay~\cite{escribano2020theoretical}.

By considering a possibility of the inclusion of a $\mathcal{B}$ boson in the VMD part of the total matrix element, we aim to reconcile the experimental data with the theoretical predictions. In our approach, we keep the L$\sigma$M part the same as described above.

\begin{figure}[h!] 
\includegraphics[width=0.5\textwidth]{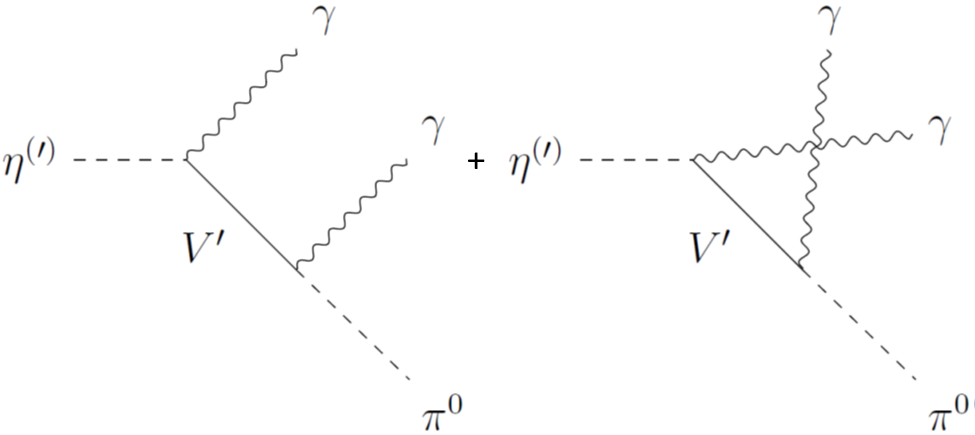}
\caption{VMD diagrams for the $\eta^{\left(\prime\right)}\rightarrow\pi^0\gamma\gamma$ decays. In addition to vector mesons, hypothetical $\mathcal{B}$ boson is taken into account, and $V^\prime = \left(\omega,\rho^0,\phi,\mathcal{B} \right)$. }
\label{Diagram}
\end{figure}

The VMD part of the total amplitude in Eqn.~(\ref{VMD}) is modified by an inclusion of the $\mathcal{B}$ boson:

\begin{equation}
    \sum_{V=\rho^0, \omega, \phi} \rightarrow \sum_{V^\prime=\rho^0
    , \omega, \phi, \mathcal{B}},
\end{equation}
with an additional element included:

\begin{equation}
c^{\mathcal{B}}_{\eta^{\left(\prime\right)}\rightarrow\pi^0\gamma\gamma}\left[\frac{(P\cdot q_2-m_{\eta^{\left(\prime\right)}}^2)\{a\}-\{b\}}{D_{\mathcal{B}}(t)}+
\bigg\{
\begin{array}{c}
q_2\leftrightarrow q_1\\
t\leftrightarrow u
\end{array}
\bigg\}\right]\
\end{equation}

In total, the VMD part of the amplitude includes 8 diagrams, including two diagrams corresponding to the hypothetical $\mathcal{B}$ boson as shown in Fig.~\ref{Diagram}.

In our considerations, we use the same propagator as in the case of vector meson, Eqn.~(\ref{BW}), $D_{\mathcal{B}}(t) = m_\mathcal{B}^2-t-i\,m_\mathcal{B}\Gamma_\mathcal{B}$, and assume $\Gamma_\mathcal{B}$ to be constant. In the next Section, we perform a fit of the coupling constants $c^{\mathcal{B}}_{\eta\rightarrow\pi^0\gamma\gamma}$ and $c^{\mathcal{B}}_{\eta^\prime\rightarrow\pi^0\gamma\gamma}$ to the experimental data using this approach. 

\section{Numerical results}\label{Numerical}

For the case of $\eta^\prime\rightarrow\pi^0\gamma\gamma$, the presence of $\mathcal{B}$ boson could be detected in the case that its peak is far enough from the $\omega$ resonance on the $\pi^0\gamma$ plot~\cite{tulin2014}. However, on an experimental $\pi^0\gamma$ plot, a sharp peak corresponding to a new particle is not observed~\cite{ablikim2017observation}. Therefore, in our calculations, we require the following condition to be fulfilled:

\begin{equation}\label{condition}
\begin{cases}
m_{\mathcal{B}} + \Gamma_{\mathcal{B}} \, \lesssim \, m_{\omega} + \Gamma_{\omega} \\ 
m_{\omega} - \Gamma_{\omega} \, \lesssim \, m_{\mathcal{B}} - \Gamma_{\mathcal{B}}
\end{cases}\Rightarrow\lvert m_{\mathcal{B}} - m_{\omega}\lvert \, \lesssim \, \Gamma_{\omega} - \Gamma_{\mathcal{B}}    
\end{equation}

For our numerical analysis, we choose the values of $m_{\mathcal{B}}$ and $\Gamma_{\mathcal{B}}$ according to the condition in Eqn.~(\ref{condition}) as: $\left(m_{\mathcal{B}} = m_{\omega} - \frac{\Gamma}{2}, 
\Gamma_{\mathcal{B}} = \frac{\Gamma_\omega}{2}\right)$, $\left( m_{\mathcal{B}} = m_{\omega},  \Gamma_{\mathcal{B}} = \frac{\Gamma_\omega}{2} \right)$, $\left(
m_{\mathcal{B}} = m_{\omega}, \Gamma_{\mathcal{B}} = \Gamma_\omega\right)$, and $\left(
m_{\mathcal{B}} = m_{\omega} + \frac{\Gamma}{2},  \Gamma_{\mathcal{B}} = \frac{\Gamma_\omega}{2}\right)$.

For these values of the parameters, we perform a fit of the values of the coupling constant in both cases,  $c^\mathcal{B}_{\eta\rightarrow\pi^0\gamma\gamma}$ and $c^\mathcal{B}_{\eta^\prime\rightarrow\pi^0\gamma\gamma}$, to minimize:

\begin{equation}
\chi^2 = \sum_k \frac{\left(\frac{\mathrm{d}\Gamma_{theory}^{\eta^{\left(\prime\right)}\rightarrow\pi^0\gamma\gamma}}{\mathrm{d}m^2_{\gamma\gamma}} - \frac{\mathrm{d}\Gamma_{experiment}^{\eta^{\left(\prime\right)}\rightarrow\pi^0\gamma\gamma}}{\mathrm{d}m^2_{\gamma\gamma}}  \right)_k^2}{\sigma_k^2}
\end{equation}

The results of the fit are presented in Table~\rm{II}. For the case of $\eta^\prime\rightarrow\pi^0\gamma\gamma$, the original VMD+L$\sigma$M already provides reasonable agreement with the experimental data, therefore the value of the fitted parameter $c^{\mathcal{B}}_{\eta^\prime\rightarrow\pi^0\gamma\gamma}$ is close to zero. However, for the case of the $\eta\rightarrow\pi^0\gamma\gamma$ decay, the value of $c^{\mathcal{B}}_{\eta\rightarrow\pi^0\gamma\gamma}$ is larger, and reconciles theoretical predictions with the experimental data. 

\begin{figure}[h!] 
\includegraphics[width=\textwidth]{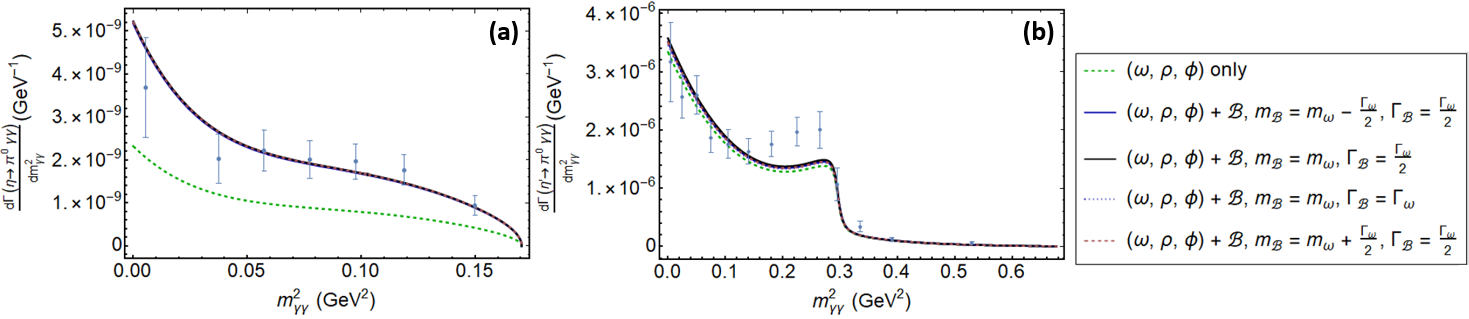}
\caption{Results on $\gamma\gamma$ spectrum from fit of the $\mathcal{B}$ boson to the experimental data from~\cite{nefkens2014new,ablikim2017observation}. The corresponding values of the coupling constant $c^{\mathcal{B}}_{\eta\rightarrow\pi^0\gamma\gamma}$ and $c^{\mathcal{B}}_{\eta^\prime\rightarrow\pi^0\gamma\gamma}$ are presented in Table~\rm{II}.}
\label{GammaGamma}
\end{figure}

For the $\gamma\gamma$ spectrum, 
an improved fit to the experimental data can be observed in Fig.~\ref{GammaGamma} (a). In Fig.~\ref{GammaGamma} (b), the corresponding fit for the $\eta^\prime\rightarrow\pi^0\gamma\gamma$ is shown. The lines corresponding to $\left(\omega,\rho^0,\phi\right)$ and $\left(\omega,\rho^0,\phi\right) + \mathcal{B}$ are close to each other, because in this case there is already a good agreement with the experimental result. 

\begin{figure}[h!] 
\includegraphics[width=0.99\textwidth]{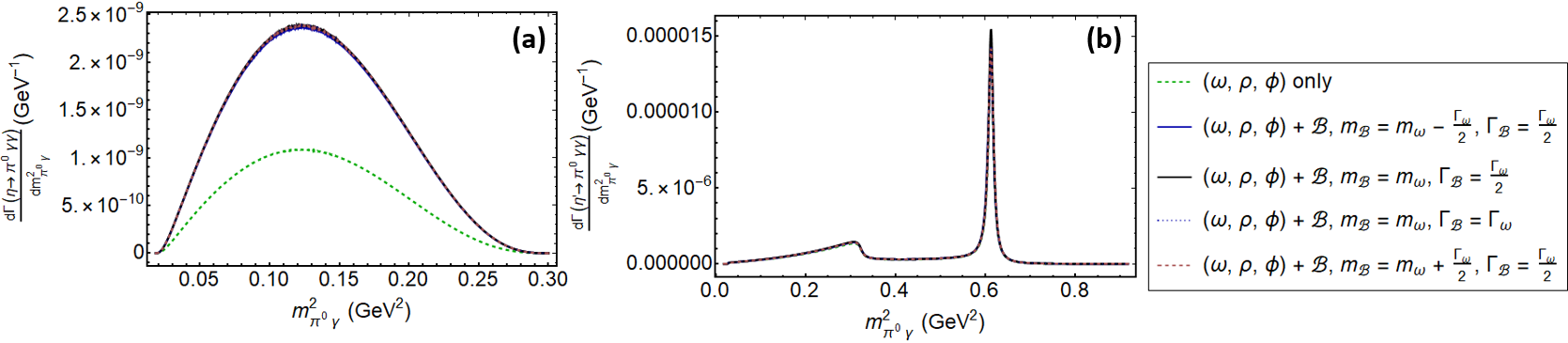}
\caption{Results on $\pi^0\gamma$ spectrum from fit of the $\mathcal{B}$ boson to the experimental data from~\cite{nefkens2014new,ablikim2017observation} The corresponding values of the coupling constant $c_{\mathcal{B}}^{\eta^\prime\rightarrow\pi^0\gamma\gamma}$ are presented in Table~\rm{II}.}
\label{PiGamma}
\end{figure}

For the $\pi^0\gamma$ spectrum, the results of our fit are provided in Fig.~\ref{PiGamma}. For the case of $\eta^\prime\rightarrow\pi^0\gamma\gamma$ shown in Fig.~\ref{PiGamma} (b), the lines corresponding to $\left(\omega,\rho^0,\phi\right)$ and $\left(\omega,\rho^0,\phi\right) + \mathcal{B}$ are almost indistinguishable, while for the case of $\eta\rightarrow\pi^0\gamma\gamma$ shown in Fig.~\ref{PiGamma}
(a), the spectrum is significantly shifted upward.

\onecolumngrid\
\twocolumngrid\

\onecolumngrid\
\twocolumngrid\

\begin{table*}
{\footnotesize{
  \centering
  \begin{tabular}{|c|l|c|c|c|c|c|c|c|c|}
\hline
    \multicolumn{1}{|c|}{Decay}& 
    \multicolumn{1}{c|}{Parameters}& 
    \multicolumn{1}{c|}{$c^{\mathcal{B}}_{\eta\rightarrow\pi^0\gamma\gamma}$ and $c^{\mathcal{B}}_{\eta^\prime\rightarrow\pi^0\gamma\gamma}$, $GeV^{-1}$}& 
    \multicolumn{1}{c|}{$\Gamma_{th}$, $GeV$}& 
    \multicolumn{1}{c|}{BR$_{th}$}&
    \multicolumn{1}{c|}{BR$_{exp}$}\\
    \hline
      \multirow{2}{*}{$\eta\to\pi^0\gamma\gamma$} & $\left(\omega,\rho^0,\phi\right)$ only  & $0$ & $1.56\left(7\right) \times 10^{-10}$ & $1.19(5)\times 10^{-4}$ &  \multirow{2}{*}{$2.56(22)\times 10^{-4}$} \\ 
& $\left(\omega,\rho^0,\phi\right) + \mathcal{B}$,  $m_{\mathcal{B}}=m_\omega - \frac{\Gamma_\omega}{2}, \Gamma_\mathcal{B} =  \frac{\Gamma_\omega}{2}$ & $0.098\left(4\right)$ & $3.41\left(5\right)\times 10^{-10}$ & $2.60\left(4\right)\times 10^{-4}$ &  
\\ 
& $\left(\omega,\rho^0,\phi\right) + \mathcal{B}$,  $m_{\mathcal{B}}=m_\omega, \Gamma_\mathcal{B} =  \frac{\Gamma_\omega}{2}$  & $0.101\left(7\right)$ & $3.43\left(8\right)\times 10^{-10}$ & $2.62(12)\times 10^{-4}$ &  
\\ 
& $\left(\omega,\rho^0,\phi\right) + \mathcal{B}$,  $m_{\mathcal{B}}=m_\omega, \Gamma_\mathcal{B} =  \Gamma_\omega$  & $0.101\left(7\right)$ & $3.44\left(7\right)\times 10^{-10}$ & $2.63(10)\times 10^{-4}$ &  
\\
& $\left(\omega,\rho^0,\phi\right) + \mathcal{B}$,  $m_{\mathcal{B}}=m_\omega+\frac{\Gamma_\omega}{2}, \Gamma_\mathcal{B} =  \frac{\Gamma_\omega}{2}$  & $0.102\left(8\right)$ & $3.43\left(7\right)\times 10^{-10}$ & $2.62(10)\times 10^{-4}$ &  
\\
\cline{5-5}
    \hline
      \multirow{2}{*}{$\eta^\prime\to\pi^0\gamma\gamma$} & $\left(\omega,\rho^0,\phi\right)$ only  & $0$ & $5.33\left(15\right) \times 10^{-7}$ & $2.81\left(9\right)\times 10^{-3}$ &  \multirow{2}{*}{$3.20(24)\times 10^{-3}$} \\ 
& $\left(\omega,\rho^0,\phi\right) + \mathcal{B}$,  $m_{\mathcal{B}}=m_\omega - \frac{\Gamma_\omega}{2}, \Gamma_\mathcal{B} =  \frac{\Gamma_\omega}{2}$ & $0.003\left(2\right)$ & $5.58\left(3\right) \times 10^{-7}$ & $2.95\left(6\right)\times 10^{-3}$ &  
\\ 
& $\left(\omega,\rho^0,\phi\right) + \mathcal{B}$,  $m_{\mathcal{B}}=m_\omega, \Gamma_\mathcal{B} =  \frac{\Gamma_\omega}{2}$  & $0.002\left(1\right)$ & $5.58\left(4\right)\times 10^{-7}$ & $2.94\left(5\right)\times 10^{-3}$ &  
\\ 
& $\left(\omega,\rho^0,\phi\right) + \mathcal{B}$,  $m_{\mathcal{B}}=m_\omega, \Gamma_\mathcal{B} =  \Gamma_\omega$  & $0.003\left(2\right)$ & $5.54\left(14\right)\times 10^{-7}$ & $2.93\left(7\right)\times 10^{-3}$ &  
\\
& $\left(\omega,\rho^0,\phi\right) + \mathcal{B}$,  $m_{\mathcal{B}}=m_\omega+\frac{\Gamma_\omega}{2}, \Gamma_\mathcal{B} =  \frac{\Gamma_\omega}{2}$  & $0.003\left(2\right)$ & $5.54\left(15\right)\times 10^{-7}$ & $2.92\left(7\right)\times 10^{-3}$ &  
\\
\cline{5-5}
   \hline
  \end{tabular}
  \label{Table}
  \caption{Results of the fit of the coupling constants of the $\mathcal{B}$ boson to the experimental data from~\cite{nefkens2014new,ablikim2017observation}.}}}
\end{table*}

Note, in both cases,  the lines corresponding to all values of the parameters were chosen according to the condition in Eqn.~(\ref{condition}) and are close to each other.

To sum up, while the impossibility of a simultaneous description of these decays may be attributed to the limitation of the VMD+L$\sigma$M framework itself, the mismatch between the  $\eta^{\left(\prime\right)}\rightarrow\pi^0\left(\eta\right)\gamma\gamma$ decays may be used as a tool to probe the parameters of the hypothetical dark photon primarily interacting with quarks rather than leptons. Even though a sharp peak on the $\pi^0\gamma$ spectrum corresponding to a new particle, which could be a definite indication of existence of the $\mathcal{B}$ boson, is not observed~\cite{ablikim2017observation}, the impossibility to reproduce the experimental data on all three decays $\eta^{\left(\prime\right)}\rightarrow\pi^0\left(\pi^0\right)\gamma\gamma$ within the VMD+L$\sigma$M framework may be an indication in favor of the leptophobic dark photon scenario. Moreover, in case precise data on all three decays becomes available, the couplings $g_{\mathcal{B}V\gamma}$ of the $\mathcal{B}$ boson can be extracted by a simultaneous fit to all three decays, as shown in Eqn.~(\ref{BCouplings}).

\section{Conclusions}\label{Conclusions}

Despite a long history of studies, both theoretically and experimentally, it seems impossible to describe the recent data on the $\eta\rightarrow\pi^0\gamma\gamma$, $\eta^\prime\rightarrow\pi^0\gamma\gamma$ and $\eta^\prime\rightarrow\eta\gamma\gamma$ decays simultaneously in the VMD+L$\sigma$M framework, even though  it is possible to reproduce the overall shapes of the spectra of these decays. 

Although meson studies have long been considered a tool for finding the dark photon, such scenarios focus primarily on lepton couplings. The possibility of a leptophobic dark photon coupled primarily to quarks rather than leptons remains an open possibility.

While the $\eta^{\left(\prime\right)}\rightarrow\pi^0\left(\eta\right)\gamma\gamma$ puzzle may be attributed to the deficiency of the VMD+L$\sigma$M, we propose to use it as a tool to fit the parameters of this hypothetical particle. There are a number of observational anomalies which can be attributed to the dark photon coupled to leptons. The $\eta^{\left(\prime\right)}\rightarrow\pi^0\left(\eta\right)\gamma\gamma$ puzzle could possibly be the first sign indicating the possibility of a dark photon coupled to quarks rather than leptons. Importantly, in order to reproduce the experimental results on the $\eta^{\left(\prime\right)}\rightarrow\pi^0\left(\eta\right)\gamma\gamma$ decays, one needs to assume ``unnatural" values of the $\mathcal{B}$ coupling constants.

With the new facilities under construction, there is a possibility for testing this kind of new physics and decisively defining whether this puzzle should be attributed to the experimental error, deficiency of the VMD+L$\sigma$M, or the leptophobic dark photon.

\section*{Code availability and reproducibility of our results}
\label{Reproducibility}

Our numerical code is publicly available on GitHub repository~\footnote{\href{https://github.com/BalytskyiJaroslaw/DarkPhoton.git}{https://github.com/BalytskyiJaroslaw/DarkPhoton.git}}{\href{https://github.com/BalytskyiJaroslaw/DarkPhoton.git}{}} to facilitate the usage of our results by other researchers.

\section*{Acknowledgments}

This work was supported, in part, by the U.S. Civilian Research \& Development Foundation (CRDF Global). Y.B. is also partially funded by the UCCS BioFrontiers institute and would like to acknowledge this funding. Y.B. appreciates useful discussions with Drs. Kelly McNear and Kyle Culhane.


\begin{thebibliography}{1}%
\makeatletter
\providecommand \@ifxundefined [1]{%
 \@ifx{#1\undefined}
}%
\providecommand \@ifnum [1]{%
 \ifnum #1\expandafter \@firstoftwo
 \else \expandafter \@secondoftwo
 \fi
}%
\providecommand \@ifx [1]{%
 \ifx #1\expandafter \@firstoftwo
 \else \expandafter \@secondoftwo
 \fi
}%
\providecommand \natexlab [1]{#1}%
\providecommand \enquote  [1]{``#1''}%
\providecommand \bibnamefont  [1]{#1}%
\providecommand \bibfnamefont [1]{#1}%
\providecommand \citenamefont [1]{#1}%
\providecommand \href@noop [0]{\@secondoftwo}%
\providecommand \href [0]{\begingroup \@sanitize@url \@href}%
\providecommand \@href[1]{\@@startlink{#1}\@@href}%
\providecommand \@@href[1]{\endgroup#1\@@endlink}%
\providecommand \@sanitize@url [0]{\catcode `\\12\catcode `\$12\catcode
  `\&12\catcode `\#12\catcode `\^12\catcode `\_12\catcode `\%12\relax}%
\providecommand \@@startlink[1]{}%
\providecommand \@@endlink[0]{}%
\providecommand \url  [0]{\begingroup\@sanitize@url \@url }%
\providecommand \@url [1]{\endgroup\@href {#1}{\urlprefix }}%
\providecommand \urlprefix  [0]{URL }%
\providecommand \Eprint [0]{\href }%
\providecommand \doibase [0]{http://dx.doi.org/}%
\providecommand \selectlanguage [0]{\@gobble}%
\providecommand \bibinfo  [0]{\@secondoftwo}%
\providecommand \bibfield  [0]{\@secondoftwo}%
\providecommand \translation [1]{[#1]}%
\providecommand \BibitemOpen [0]{}%
\providecommand \bibitemStop [0]{}%
\providecommand \bibitemNoStop [0]{.\EOS\space}%
\providecommand \EOS [0]{\spacefactor3000\relax}%
\providecommand \BibitemShut  [1]{\csname bibitem#1\endcsname}%
\let\auto@bib@innerbib\@empty
\bibitem [{Note1()}]{Note1}%
  \BibitemOpen
  \bibinfo {note} {\protect \href
  {https://github.com/BalytskyiJaroslaw/DarkPhoton.git}{https://github.com/BalytskyiJaroslaw/DarkPhoton.git}}\BibitemShut
  {NoStop}%
\end{thebibliography}%


\begin{thebibliography}{99}

\bibitem{Gan:2020aco} 
L.~Gan, B.~Kubis, E.~Passemar and S.~Tulin, Phys.\ Rept.\  {\bf 945}, (2022): 1 - 105.

\bibitem{Landsberg:1985}
L.~G.~Landsberg, Phys.\ Rept.\  {\bf 128}, 301 (1985).

\bibitem{Alde1984}
D.~Alde {\it et al.}  [Serpukhov-Brussels-Annecy(LAPP) and Soviet-CERN Collaborations], Z.\ Phys.\ C {\bf 25}, 225 (1984) [Yad.\ Fiz.\  {\bf 40}, 1447 (1984)].

\bibitem{Prakhov2005}
S.~Prakhov {\it et al.},
Phys.\ Rev.\ C {\bf 72}, 025201 (2005).

\bibitem{Prakhov2008}
S.~Prakhov {\it et al.}, Phys.\ Rev.\ C {\bf 78}, 015206 (2008).

\bibitem{Knecht2004}
N.~Knecht {\it et al.}, Phys.\ Lett.\ B {\bf 589}, 14 (2004).


\bibitem{dimillo2006}
B.~Di Micco {\it et al.} [KLOE Collaboration], Acta Phys.\ Slov.\  {\bf 56}, 403 (2006).

\bibitem{nefkens2014new}
B.~M.~K.~Nefkens {\it et al.} [A2 at MAMI Collaboration], Phys.\ Rev.\ C {\bf 90}, 025206 (2014).

\bibitem{PDG}
P.A. Zyla \textit{et al.} (Particle Data Group), Prog. Theor. Exp. Phys. {\bf 2020}, 083C01 (2020) and 2021 update.

\bibitem{ablikim2017observation} 
M.~Ablikim {\it et al.} [BESIII Collaboration], Phys.\ Rev.\ D {\bf 96}, 012005 (2017).

\bibitem{alde1987neutral} 
D.~Alde {\it et al.}  [Serpukhov-Brussels-Los Alamos-Annecy(LAPP) Collaboration], Z.\ Phys.\ C {\bf 36}, 603 (1987).

\bibitem{ablikim2019search} 
M.~Ablikim {\it et al.} [BESIII Collaboration], Phys.\ Rev.\ D {\bf 100}, 052015 (2019).

\bibitem{JLab1}
L.  Gan  {\it et  al.}, JLab  proposal, \href{https://www.jlab.org/exp_prog/proposals/14/PR12-14-004.pdf}{link}.

\bibitem{JLab2}
L.~P.~Gan and A.~Gasparian, PoS CD {\bf 09}, 048 (2009).

\bibitem{JLab3} L.~Gan, PoS CD {\bf 15}, 017 (2015).

\bibitem{JLab4} L.~Gan {\it et al.}, Update to the JEF proposal, JLab proposal, \href{https://www.jlab.org/exp_prog/proposals/17/C12-14-004.pdf}{link}.

\bibitem{JLab5} D.~Lawrence [GlueX Collaboration], AIP Conf.\ Proc.\ {\bf 1182}, 811 (2009).

\bibitem{oppo1967models} 
G.~Oppo and S.~Oneda,
Phys.\ Rev.\  {\bf 160}, 1397 (1967).

\bibitem{baracca1970general} 
A.~Baracca and A.~Bramon,
Nuovo Cim.\ A {\bf 69}, 613 (1970).

\bibitem{ametller1992chiral} 
L.~Ametller, J.~Bijnens, A.~Bramon and F.~Cornet,
Phys.\ Lett.\ B {\bf 276}, 185 (1992).

\bibitem{ko1993contributions} 
P.~Ko, Phys.\ Rev.\ D {\bf 47}, 3933 (1993).

\bibitem{ko1995eta} 
P.~Ko, Phys.\ Lett.\ B {\bf 349}, 555 (1995).

\bibitem{oset2003eta} 
E.~Oset, J.~R.~Pelaez and L.~Roca, Phys.\ Rev.\ D {\bf 67}, 073013 (2003).

\bibitem{oset2008eta} 
E.~Oset, J.~R.~Pelaez and L.~Roca, Phys.\ Rev.\ D {\bf 77}, 073001 (2008).

\bibitem{danilkin2017theoretical} 
I.~Danilkin, O.~Deineka and M.~Vanderhaeghen,
  Phys.\ Rev.\ D {\bf 96}, 114018 (2017).
  
\bibitem{belkov1995} 
A.~A.~Bel'kov, A.~V.~Lanyov and S.~Scherer,
J.\ Phys.\ G {\bf 22}, 1383 (1996).

\bibitem{bellucci1995} 
S.~Bellucci and C.~Bruno, Nucl.\ Phys.\ B {\bf 452}, 626 (1995).

\bibitem{bijnens1995} 
J.~Bijnens, A.~Fayyazuddin and J.~Prades,
Phys.\ Lett.\ B {\bf 379}, 209 (1996).

\bibitem{ng1993} 
J.~N.~Ng and D.~J.~Peters,
Phys.\ Rev.\ D {\bf 47}, 4939 (1993).

\bibitem{nemoto1996} 
Y.~Nemoto, M.~Oka and M.~Takizawa, Phys.\ Rev.\ D {\bf 54}, 6777 (1996).

\bibitem{escribano2012}
R.~Escribano, PoS QNP {\bf 2012}, 079 (2012).

\bibitem{jora2010} 
R.~Jora, Nucl.\ Phys.\ Proc.\ Suppl.\  {\bf 207-208}, 224 (2010).

\bibitem{Balytskyi:2018pzb} 
Y.~Balytskyi, arXiv:1804.02607 [hep-ph].

\bibitem{Balytskyi:2018uxb} 
Y.~Balytskyi,  LHEP-{\bf 156}, (2020).

\bibitem{escribano2020theoretical}
R.~Escribano, S.~Gonzalez-Solis, R.~Jora, and E.~Royo, Phys.\ Rev.\ D {\bf 102}, 034026 (2020).

\bibitem{Bramon:2000fr}
A.~Bramon, R.~Escribano and M.~Scadron,
Phys.\ Lett.\ B {\bf 503} (2001), 271-276.

\bibitem{Escribano:2020jdy}
R.~Escribano and E.~Royo, Phys.\ Lett.\ B {\bf 807} (2020), 135534.

\bibitem{nelson1989}
A.~E.~Nelson and N.~Tetradis, Phys.\ Lett.\ B {\bf 221}, 80 (1989).

\bibitem{fayet2006}
P.~Fayet, Phys.\ Rev.\ D {\bf 74}, 054034 (2006).

\bibitem{arkani2009}
N.~Arkani-Hamed, D.~P.~Finkbeiner, T.~R.~Slatyer and N.~Weiner, Phys.\ Rev.\ D {\bf 79}, 015014 (2009).

\bibitem{pospelov2009}
M.~Pospelov and A.~Ritz, Phys.\ Lett.\ B {\bf 671}, 391 (2009).

\bibitem{tulin2018}
S.~Tulin and H.~B.~Yu, Phys.\ Rept.\ {\bf 730}, 1 (2018).

\bibitem{Aoyama2020}
T.~Aoyama {\it et al.}, Phys. \ Rep. \ {\bf 887}, 1 (2020).

\bibitem{Brookhaven}
G.~W.~Bennett {\it et al.}, [Muon g-2 Collaboration], Phys. \ Rev. \ D {\bf 73}, 072003 (2006).

\bibitem{FermiLab1}
B.~Abi {\it et al.}, [Muon g-2 Collaboration], Phys. \ Rev. \ Lett. {\bf 126},  141801 (2021).

\bibitem{FermiLab2}
T.~Albahri {\it et al.}, [Muon g-2 Collaboration], Phys. \ Rev. \ D {\bf 103}, 072002 (2021).

\bibitem{FermiLab3}
J.~Grange {\it et al.}, [Muon g-2 Collaboration], arXiv:1501.06858 [physics.ins-det].

\bibitem{JPARC}
N.~Saito, [J-PARC g-2/EDM Collaboration], AIP Conf. \ Proc. \ 1467, 45  (2012).


\bibitem{fayet2007}
P.~Fayet, Phys.\ Rev.\ D {\bf 75}, 115017 (2007).

\bibitem{pospelov20091}
M.~Pospelov, Phys.\ Rev.\ D {\bf 80}, 095002 (2009).

\bibitem{Be1}
F.~W.~N.~de~Boer {\it et al.}, Phys. \ Lett. \ B {\bf 388}, 235 (1996).

\bibitem{Be2}
F.~W.~N.~de Boer, R.~van~Dantzig, J.~van~Klinken, K.~Bethge,
H.~Bokemeyer, A.~Buda, K.~A. ~M{\"u}ller, and K.~E.~Stiebing,
J. \ Phys. \ G {\bf 23}, L85 (1997).

\bibitem{Be3}
F.~W.~N.~de~Boer, K.~Bethge, H.~Bokemeyer, R.~van~Dantzig,
J.~van~Klinken, V.~Mironov, K.~A. M{\"u}ller, and K.~E.~Stiebing, J. \ Phys. \ G {\bf 27}, L29 (2001).

\bibitem{Be4}
F.~W.~N.~de~Boer, K.~Bethge, H.~Bokemeyer, R.~van~Dantzig,
J.~van~Klinken, V.~Mironov, K.~A. M{\"u}ller, and K.~E.~Stiebing, J. \ Phys. \ G {\bf 27}, L29 (2001).

\bibitem{Be5}
A.~Cs.~Vit{\'e}z {\it et al.}, Acta \ Phys. \ Pol. \ B {\bf 39}, 483 (2008).

\bibitem{Be6}
A.~Krasznahorkay {\it et al.}, Frascati \ Physics \ Series {\bf 56}, 86 (2013).

\bibitem{Be7}
D.~R.~Tilley, J.~H.~Kelley, J.~L.~Godwin, D.~J.~Millener, J.~E.~Purcell, C.~G.~Sheu, and H.~R.~Weller, Nucl. \ Phys. \ {\bf A 745}, 155 (2004).

\bibitem{krasznahorkay2016}
A.~J.~Krasznahorkay {\it et al.}, Phys.\ Rev.\ Lett.\ {\bf 116}, 042501 (2016)

\bibitem{lee1955}
T.~D.~Lee and C.~N.~Yang, Phys.\ Rev.\ {\bf 98}, 1501 (1955).

\bibitem{pais1973}
A.~Pais, Phys.\ Rev.\ D {\bf 8}, 1844 (1973).

\bibitem{rajpoot1989}
S.~Rajpoot, Phys.\ Rev.\ D {\bf 40}, 2421 (1989).

\bibitem{foot1989}
R.~Foot, G.~C.~Joshi, H.~Lew, Phys. \ Rev. \ D {\bf 40} 2487, (1989).

\bibitem{he1990}
X.~G.~He and S.~Rajpoot, Phys.\ Rev.\ D {\bf 41}, 1636 (1990).

\bibitem{carone1995}
C.~D.~Carone and H.~Murayama, Phys.\ Rev.\ Lett.\ {\bf 74}, 3122 (1995).

\bibitem{bailey1995}
D.~C.~Bailey and S.~Davidson, Phys.\ Lett.\ B {\bf 348}, 185 (1995).

\bibitem{carone19951}
C.~D.~Carone and H.~Murayama, Phys.\ Rev.\ D {\bf 52}, 484 (1995)

\bibitem{aranda1998}
A.~Aranda and C.~D.~Carone, Phys.\ Lett.\ B {\bf 443}, 352 (1998).

\bibitem{perez2010}
P.~Fileviez Perez and M.~B.~Wise, Phys.\ Rev.\ D {\bf 82}, 011901 (2010) [Erratum: Phys.\ Rev.\ D {\bf 82}, 079901 (2010)].

\bibitem{agashe2004}
K.~Agashe, G.~Servant, Phys. \ Rev. \ Lett. {\bf 93} 231805, (2004).

\bibitem{farrar2006}
G.~R.~Farrar, G.~Zaharijas, Phys. \ Rev. \ Lett. {\bf 96} 041302,  (2006).

\bibitem{davoudiasl2010}
H.~Davoudiasl, D.~E.~Morrissey, K.~Sigurdson, S.~Tulin, Phys. \ Rev. \ Lett.  {\bf 105}  211304, (2010).

\bibitem{graesser}
M.~L.~Graesser, I.~M.~Shoemaker, L.~Vecchi, arXiv:1107.2666 [hep-ph].

\bibitem{duerr2014}
M.~Duerr, P.~Fileviez P{\'e}rez, Phys. \ Lett. \ B {\bf 732} 101,  (2014).

\bibitem{nussinov1985}
S.~Nussinov, Phys. \ Lett. \ {\bf 165B}  55, (1985).

\bibitem{duerr2018}
M.~Duerr, K.~Schmidt-Hoberg, J.~Unwin, Phys. \ Lett. \ B {\bf 780} 553, (2018).

\bibitem{barbieri1975}
R.~Barbieri, T.~E.~O.~Ericson, Phys. \ Lett. \ {\bf 57B} 270, (1975).

\bibitem{leeb1992}
H.~Leeb, J.~Schmiedmayer, Phys. \ Rev. \ Lett. {\bf 68} 1472, (1992). 

\bibitem{nesvizhevsky2008}
V.~V.~Nesvizhevsky, G.~Pignol, K.~V.~Protasov, Phys. \ Rev. \ D {\bf 77}  034020, (2008).

\bibitem{adelberger2003}
E.~G.~Adelberger, B.~R.~Heckel, A.~E.~Nelson, Ann. \ Rev. \ Nucl. \ Part. \ Sci. {\bf 53} 77, (2003).

\bibitem{barger1996}
V.~D.~Barger, K.~-m.~Cheung, P.~Langacker, Phys. \ Lett. \ B {\bf 381} 226, (1996). 

\bibitem{dobrescu2013}
B.~A.~Dobrescu, F.~Yu, Phys. \ Rev. \ D {\bf 88} 035021, (2013)  [Erratum: Phys.\ Rev.\ D {\bf 90} (2014) 079901].

\bibitem{batell2014}
B.~Batell, P.~deNiverville, D.~McKeen, M.~Pospelov, A.~Ritz, Phys. \ Rev. \ D {\bf 90} 115014, (2014).

\bibitem{cohen2015}
T.~Cohen, M.~Lisanti, H.~K.~Lou, Phys. \ Rev. \ Lett. {\bf 115} 171804,  (2015).

\bibitem{berryman2021}
J.~M.~Berryman, S.~Gardner,  Phys.\ Rev.\ C {\bf 104}, 045802 (2021).

\bibitem{berryman2022}
J.~M.~Berrymana, S.~Gardner, M.~Zakeric, INT-PUB-22-001, N3AS-22-004, 	arXiv:2201.02637.

\bibitem{tulin2014}
S.~Tulin, Phys.\ Rev.\ D {\bf 89}, 114008 (2014).

\bibitem{sakurai1960}
J.~J.~Sakurai, Ann. \ Phys. \ (N.Y.) {\bf 11}, 1 (1960).

\bibitem{kroll1960}
N.~M.~Kroll, T.~D.~Lee, and B.~Zumino, Phys. \ Rev. \ {\bf 157}, 1376 (1967).

\bibitem{bando1985}
M.~Bando, T.~Kugo, S.~Uehara, K.~Yamawaki, and T.~Yanagida, Phys. \ Rev. \ Lett. \ {\bf 54}, 1215 (1985).

\bibitem{bando19851}
M.~Bando, T.~Kugo, and K.~Yamawaki, Nucl. \ Phys. \ {\bf B259},
493 (1985).

\bibitem{bando1988}
M.~Bando, T.~Kugo, and K.~Yamawaki, Phys. \ Rep. \ {\bf 164}, 217
(1988).

\bibitem{fujiwara1985}
T.~Fujiwara, T.~Kugo, H.~Terao, S.~Uehara, and K.~Yamawaki, Prog. \ Theor. \ Phys. \ {\bf 73}, 926 (1985).

\bibitem{ivanov2008}
I.~P.~Ivanov, S.~Pacetti, Eur.\ Phys. \ J. C {\bf 53}, 559 - 566 (2008).


\bibitem{du2020}
M.-L.~Du, V.~Baru, F.-K.~Guo, C. ~Hanhart, U.-G.~Meissner, A.~Nefediev, I.~Strakovsky, Eur. \ Phys. \ J. \ C {\bf 80}(11), 1053 (2020).

\bibitem{xu2021}
Y.-Z.~Xu, S,-Y.~Chen, Z.-Q.~Yao, D.~Binosi, Z.-F.~Cui, C.~D.~Roberts,
Eur. \ Phys. \ J. \ C 81:895, (2021).

\bibitem{donoghue1989}
J.~F.~Donoghue, C.~Ramirez and G.~Valencia, Phys. \ Rev. \ D {\bf 39}, 1947 (1989).

\bibitem{williams2011}
M.~Williams, C.~Burgess, A.~Maharana, and F.~Quevedo,
J. \ High \ Energy \ Phys. {\bf 08} 106, (2011).

\bibitem{babar2014}
J.~P.~Lees, Phys. \ Rev. \ Lett. {\bf 113}, 201801 (2014).

\bibitem{babar2016}
J.~P.~Lees {\it et al.}, Phys. \ Rev. \ D {\bf 94}, 011102(R) (2016).

\bibitem{babar2017}
J.~P.~Lees {\it et al.}, Phys. \ Rev. \ Lett. {\bf 119}, 131804 (2017).

\bibitem{babar2022}
J.~P.~Lees {\it et al.}, Phys. \ Rev. \ Lett. {\bf 128}, 021802 (2022).

\bibitem{kloe2012}
F.~Archilli {\it et al.}, Phys. \ Lett. \ B {\bf 706} 251-255,  (2012).

\bibitem{kloe2016}
F.~Archilli {\it et al.}, Phys. \ Lett. \ B {\bf 757} 356-361,  (2016).

\bibitem{kloe2018}
F.~Archilli {\it et al.}, Phys. \ Lett. \ B {\bf 784} 336-341,  (2018).

\bibitem{altarelli2014}
G.~Altarelli, RM3-TH/13-7; CERN-PH-TH/2013-182, arXiv:1308.0545v2, (2014).

\bibitem{hossenfelder2021}
S.~Hossenfelder, Synthese 198.16: 3727-3745, (2021).

\bibitem{bali2021}
G.~S.~Bali, V.~Braun, S.~Collins, A.~Sch{\" a}fer, and Jakob Simeth, J. \ High \ Energy \ Phys. {\bf 08} 137, (2021).

\bibitem{weinberg1979}
S.~Weinberg, Physica \ A \ {\bf 96} 327, (1979).

\bibitem{gasser1984}
J.~Gasser, H.~Leutwyler, Ann. \ Physics \ {\bf 158}  142, (1984).

\bibitem{gasser1985}
J.~Gasser, H.~Leutwyler, Nucl. \ Phys. \ B {\bf 250} 465, (1985).

\bibitem{RChTReview}
J.~Portoles, in AIP Conference Proceedings (Vol. 1322, No. 1, pp. 178-187), American Institute of Physics (2010).

\bibitem{escribano2006}
R.~Escribano, Phys. \ Rev. \ D {\bf 74}, 114020 (2006).

\bibitem{lichard2006}
P.~Lichard, M.~Vojik, [arXiv:0611163v1[hep-ph]].

\bibitem{roos1969}
M.~Roos, J.~Pisut, Nucl.\ Phys.\ B {\bf 10}, 8.B.6 (1969). 

\bibitem{bramon1995}
A.~Bramon, A.~Grau, G.~Pancheri, Phys. \ Lett. \ B {\bf 344},240-244 (1995).

\bibitem{bramon1997}
A.~Bramon, R.~Escribano, and M.~D.~Scadron, Phys.\ Lett. \ B {\bf 403}, 339 (1997).

\end{thebibliography}
\end{document}